\documentstyle[epsf,12pt]{article}
\def \gama5       {\gamma^{5}}
\def \gamaa       {\gamma^{\alpha}}

\def \ubar        {\overline{u}}
\def \Ubar        {\overline{U}}
\def \sigmaab     {\sigma_{\alpha\beta}}
\def \fplus       {f_{+}}
\def \fminus      {f_{-}}

\def \fs          {f_{S}}
\def \ft          {f_{T}}
\def \Pl          {p_{\ell}}
\def \Pnu         {p_{\nu}}

\def \Em          {E_{m}}
\def \El          {E_{\ell}}
\def \Enu         {E_{\nu}}
\def \ml          {m_{\ell}}
\def \mpi         {m_{\pi}}
\def \mh          {M_{H^{\pm}}}
\def \Tpi         {T_{\pi}}

\def \half        {\frac{1}{2}}
\def \quarter     {\frac{1}{4}}
\def \3quarter    {\frac{3}{4}}
\begin{document}
\title{Charged Higgs and Scalar Couplings in Semileptonic Meson Decay}
\author{Richard J. Tesarek}
%
%
\maketitle
\vspace{-3.2in}
\begin{flushright}
  RUTGERS-99-04\\
\end{flushright}
\vspace{2.1in}
\begin{center}
 {\it Department of Physics and Astronomy, 
      Rutgers University \\
      Piscataway, New Jersey 08855\\\vspace{10pt} 
      {\tt tesarek@physics.rutgers.edu}\\ \medskip}
 \medskip
 \begin{abstract}
  We present a new charged Higgs search technique using the effects of 
  scalar dynamics in semileptonic meson decay.  Applying this method 
  to a modest sample of $B$ meson decays yields sensitivity to the 
  high $\tan{\beta}$ region well beyond existing charged Higgs searches.
 \end{abstract}
\end{center}

Past searches for charged Higgs can be put into four categories:
1) direct production in colliders, 2) measurement of anomalously large 
branching ratios for top, bottom and $\tau$ lepton decays, 
3) lepton polarization measurements in meson decay and 4) precision 
measurements of well known quantities such as $K^0$--$\overline{K^0}$ 
mixing, $Z$ width, etc.~\cite{chhiggs:review}.  Category~1 represent
direct searches while categories 2--4 are indirect search techniques.
Of the indirect searches, only the lepton polarization technique makes 
use of the scalar nature of the charged Higgs.  Requiring scalar 
dynamics in a process which may be mediated by a charged Higgs provides 
an extra constraint and narrows the number of possible interpretations 
of indirect searches.

In this Letter, we review a method to identify scalar dynamics or 
couplings in semileptonic decays of pseudoscalar mesons.  By identifying 
the scalar coupling as a charged Higgs mediating the decay, one may 
extract the relative $H^{\pm}/W^{\pm}$ coupling strengths.  We apply 
the relative coupling strength information to a two Higgs doublet 
model and evaluate the possible sensitivities of precision measurements 
in $B$ meson decay dynamics to charged Higgs searches.

\section{General Phenomenology}
We begin by considering a general semileptonic meson decay.
\begin{equation}
  M\rightarrow m\ell\nu,
  \label{decay}
\end{equation}
where $M$ and $m$ denote the parent and offspring pseudoscalar mesons 
and $\ell$ and $\nu$ refer to the charged lepton and its neutrino.  The
general amplitude describing a pseudoscalar to pseudoscalar transition,
consistent with the Dirac equation and left handed, massless neutrinos, is
\begin{equation}
  \label{matrix_element}
  \begin{array}{cc}
  {\cal M} = &\frac{G_F}{\sqrt{2}}V_{ij}\ubar(\Pnu)(1+\gamma^5) 
              \left\{M\fs
               -\frac{1}{2}\left[(P + p)_{\alpha}\fplus \right.\right.\\
             & + \left.\left.     (P - p)_{\alpha}\fminus \right] \gamaa 
               + i \frac{\ft}M\sigmaab P^{\alpha}p^{\beta}
          \right\} v(\Pl),
  \end{array}
\end{equation}
where $G_{F}$ is the Fermi coupling constant, $V_{ij}$ is the 
appropriate Cabibbo-Kobayashi-Maskawa~\cite{CKM_matrix} (CKM) matrix 
element and $P$, $M$ and $p$, $m$ are the 4-momenta and masses of the 
parent and offspring mesons, respectively.  This transition amplitude
contains four form factors, $f_{S}$, $f_{+}$, $f_{-}$, and $f_{T}$, 
which parameterize the $M\rightarrow m$ transition and 
provide a measure of the admixture of different dynamics or couplings 
occurring in the decay.  In general, the form factors depend on the 
4-momentum transferred to the leptons, $Q^2 = (P-p)^2$.  Two of the form 
factors, $\fplus$ and $\fminus$, arise from a vector particle mediating 
the decay while the remaining form factors, $\fs$ and $\ft$, come from 
scalar and tensor exchange.  The term in equation~\ref{matrix_element} 
involving $\fminus$ may be collapsed, using the Dirac equation, to give 
an induced scalar coupling.
The tensor term may be similarly collapsed into induced vector and 
scalar components.
  
To calculate the decay rate, we use the notation of 
Chizhov~\cite{ke3th:chizhov} and define parameters for effective 
vector and scalar terms:
\begin{equation}
 \begin{array}{ll}
   V =& \fplus + \frac{\ml}M\ft \hfill\\
   S =& \fs + \frac{\ml}{2M}\fminus +  
           \left\{ \frac{(E_{\nu} - E_{\ell})}{M} +\frac{m_{\ell}^2}{2M^2} 
          \right\}\ft.
  \end{array}
  \label{dk_param}
\end{equation}
The decay rate can then be calculated
in the rest frame of the parent meson.
\begin{equation}
  \Gamma(\Em , \El) \propto A\cdot |V|^2 + B\cdot\Re(V^*S) + C\cdot |S|^2,
  \label{dk_rate}
\end{equation}
where $A$, $B$ and $C$ depend on the kinematics of the decay.
\begin{eqnarray}
  A &=& M(2\El\Enu - M\Delta\Em) 
    -\ml^{2}\left(\Enu - \quarter\Delta\Em\right),\nonumber\\ 
  B &=& mM (2\Enu - \Delta\Em), \\
  C &=& M^{2}\Delta\Em,\nonumber
\end{eqnarray}
and
\begin{equation}
  \Delta\Em = \frac{M^2+m^2-\ml^2}{2M} -\Em.
\end{equation}
From equation~\ref{dk_param}, note that the induced scalar coupling 
is suppressed for heavy mesons decaying into light leptons by the 
factor $\ml/M$.  This suppression factor limits the ability to search 
for scalar effects unless one has {\it a priori} knowledge of $\fminus$.  
This limitation will be discussed when applying the general analysis to 
specific decays.

As an aside, we observe that the form factors may be complex with 
non-zero phases.  While an overall phase is unobservable, the decay 
rate is sensitive to the relative phases of the form factors,
$\phi_{ij} = \phi_i - \phi_j,$ where $i$, $j$ denote the form factor 
indices:  $+$, $-$, $S$ or $T$.  The interference terms between the 
form factors then enter the decay rate as:
\begin{equation}
  \Re{\{f_i^*f_j\}} = |f_i(Q^2)||f_j(Q^2)|\cos{\phi_{ij}}.
\end{equation}
It is interesting to note that if $\fplus$, $\fminus$, $\fs$ and $\ft$
are not all relatively real ($\phi_{ij} = 0,\pi$) then CP is violated.

The scalar and tensor coupling strengths may be isolated by 
reparameterizing these form factors as a product of a structure 
dependent term which depends on the momentum transfer, $F_i(Q^2)$, 
and a relative coupling strength, $g_i$;
\begin{equation}
  f_i(Q^2) = F_i(Q^2)g_i,
\end{equation}
where $i$ denotes either $S$ or $T$.  In order to be sensitive to 
deviations from pure vector behavior, we remove a common factor of 
$\fplus(0)$ from equation~\ref{matrix_element}.
Then, in the limit of no momentum transferred to the leptons ($Q^2=0$),
\begin{equation}
  \left|\frac{f_i(0)}{\fplus(0)}\right| = 
          \frac{|F_i|}{|\fplus|} g_i,
          \label{ffratio}
\end{equation}
where the explicit reference to the $Q^2$ dependence is dropped.
For the remainder of this Letter, form factors will be assumed to be 
evaluated at $Q^2=0$.  In order to extract the relative coupling 
strengths, $g_i$, from a form factor ratio, one need only 
know the ratio $|F_i/\fplus|$.  For tensor couplings, this ratio
must be evaluated on a case-by-case basis.  Note that in the
heavy quark limit, $|\fminus/\fplus| = 1$~\cite{HQET}.
Since the induced scalar is indistinguishable
from a true scalar exchange, we argue that $|F_S/\fplus| =1$.

\begin{figure}[tbp]
%
%
  \epsfxsize=5.0in \epsfysize=6.4in
  \epsffile{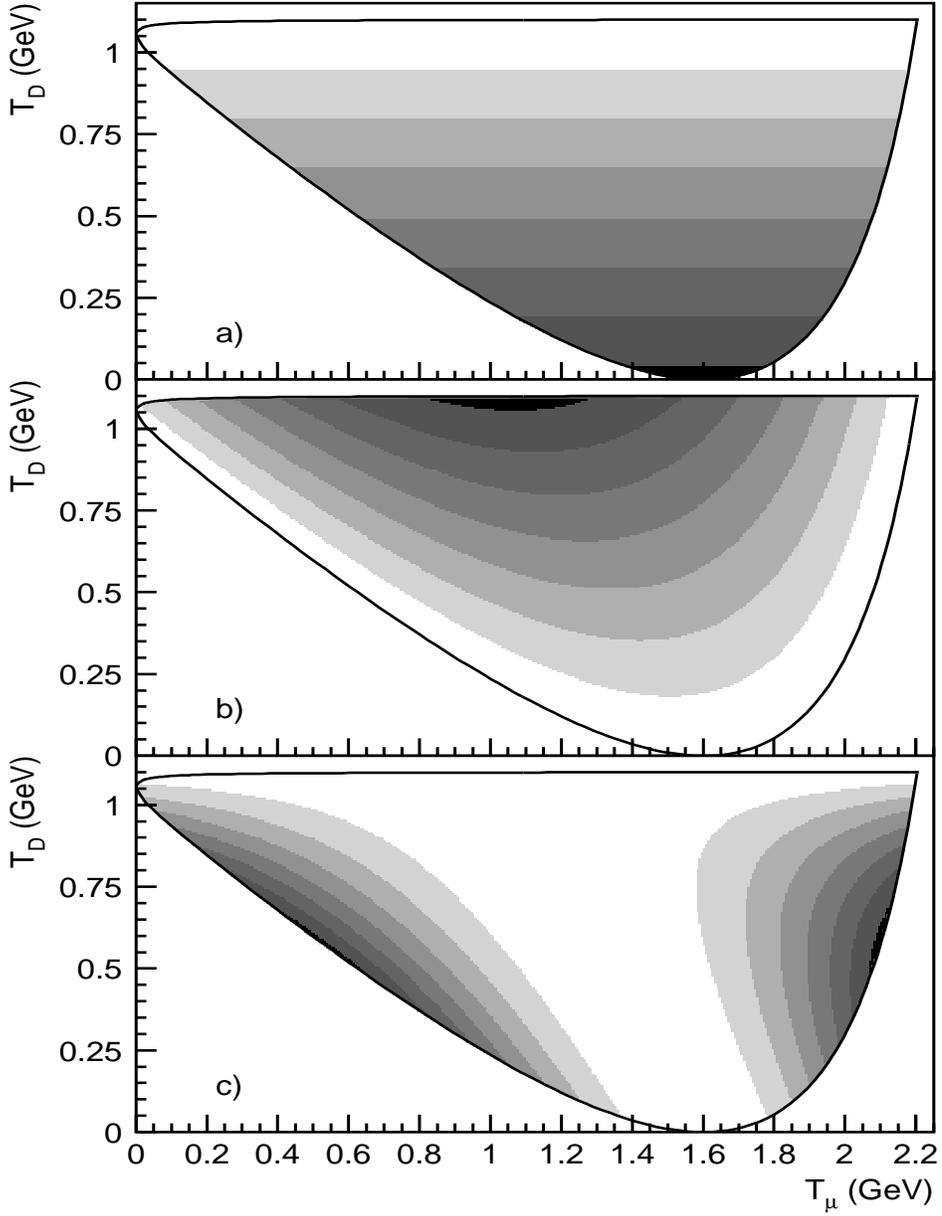}
  \caption{\label{bmu3_dynamics}  Phase space density contours for 
   decays of $B^0\rightarrow D^- \mu^+\nu$ showing the effects of 
   a) scalar, b) vector and c) tensor couplings in the decay.  
   The variables plotted are the $D^-$ and $\mu^+$ kinetic energies 
   $(T_{D},T_{\mu})$.   Darker colors represent a higher phase space
   plot density.}
\end{figure}

Any two Lorentz invariants may be used to describe the phase space
of a three body decay.  For semileptonic decays, it is traditional
to choose the kinetic energies of the offspring meson 
($T_m = E_m - m_m$) and the lepton ($T_{\ell} = E_{\ell} - m_{\ell}$) 
where both energies are measured in the parent meson rest frame.
Figure~\ref{bmu3_dynamics} shows how the different couplings affect 
the phase space density or Dalitz plot for $B^0\rightarrow D^- \mu^+ \nu$ 
decay.  The figure shows that the character of the Dalitz plot changes 
dramatically depending on the Lorentz structure of the coupling.  

Analysis of the shape of the Dalitz plot for semileptonic decays
would then yield information about possible tensor and scalar couplings.
Non-zero values of either tensor or scalar couplings would indicate 
the onset of new types of physics not predicted by the Standard Model 
of particle physics. Tensor couplings have previously been discussed 
in the literature~\cite{tensor:summary,tensor:poblaguev} and are beyond 
the scope of this Letter.  Scalar couplings have also been discussed in 
the context of charged Higgs exchange~\cite{tensor:poblaguev}.  Outside 
of CP violating effects, most of the charged Higgs phenomenology is 
discussed in terms of heavy quark or $\tau$ decay~\cite{chhiggs:review} 
where the couplings are expected to be the strongest.  However, from 
equation~\ref{dk_rate} the decays of mesons into relatively light 
($\ml/M \ll 1$) leptons provide an interesting probe for scalar effects 
and may yield information on charged Higgs.  In this context, we now 
evaluate charged Higgs couplings.
\section{Charged Higgs Couplings}
In theories where multiple Higgs doublets are responsible for 
the spontaneous symmetry breaking of the electromagnetic and weak
interactions, charged Higgs particles arise as a consequence of the
theory.  A charged Higgs particle may mediate semileptonic decays 
in the same manner as a $W^{\pm}$.  The differences between 
$H^{\pm}$ and $W^{\pm}$ mediation arise from the Lorentz structure 
of the coupling and the coupling strength.  
In general, the Higgs-fermion coupling strengths are model dependent.  
We consider a type-II two Higgs doublet model where one Higgs doublet, 
$\phi_2$, couples to down-type quarks and charged leptons and the 
other Higgs doublet couples to up-type quarks.  Minimal supersymmetry predicts 
a type-II Higgs model with added constraints.  For a large charged 
Higgs mass ($\mh\gg M$) one may write down the Lagrangian density 
describing the Higgs-fermion interaction using the Fermi formalism.  
In terms of the ratio of vacuum expectation values 
$X=v_2/v_1\equiv\tan{\beta}$ and $Y=1/X\equiv\cot{\beta}$ 
the Lagrangian density for type-II Higgs models is~\cite{chhiggs-II}:
\begin{eqnarray}
  {\cal L}_H  &=& 2^{\3quarter }G_F^{\half}H^+ \left\{ \Ubar
     \left[ X V_{ij}M_D(1 + \gama5 )
     +  Y M_UV_{ij}(1 - \gama5 ) \right] D \right. \nonumber\\
   && \left. + \overline\nu  X M_L(1 + \gama5 ) L \right\} + {\rm h.c.},
\end{eqnarray}
where $M_D$, $M_U$ and $M_L$ are the diagonal mass matrices, and
\begin{equation}
  U = \left( \begin{array}{c} u \\ c \\ t \end{array} \right), 
  D = \left( \begin{array}{c} d \\ s \\ b \end{array} \right), 
  \\
  L = \left( \begin{array}{c} e \\ \mu \\ \tau \end{array} \right), 
  \nu = \left( \begin{array}{c} \nu_e \\ \nu_{\mu} \\ \nu_{\tau} 
  \end{array} \right).
\end{equation}
For the theory to be perturbative regime and allow us to write and
evaluate Feynman diagrams for the decay processes, the $H^+$--$t$,
$H^+$--$b$ coupling strength must be small. This leads to two 
limiting conditions:
\begin{equation}
   \begin{array}{ccc}
    X \ll 1,  & \frac{G_F}{8\pi\sqrt{2}}m_b^2 X^2 & \ll 1, \\
    X \gg 1,  & \frac{G_F}{8\pi\sqrt{2}}m_t^2 Y^2 & \ll 1,
   \end{array} 
\end{equation}
which reduces to the range $0.20 < X < 200$.

The quark level transition amplitudes for the decay in equation~\ref{decay} 
can then be written for mesons containing down-type quarks decaying into
mesons containing up-type quarks (eg. $s\rightarrow u\ell\nu$, 
$b\rightarrow c\ell\nu$, etc).
\begin{eqnarray}
  {\cal A}_H &=i\frac{G_F}{\sqrt{2}}\frac{V_{ij}}{\mh^2} 
     \left\{\overline{U} \left[ X m_D(1 + \gama5 ) 
        + Y m_U(1 - \gama5 )\right] D\right\} \nonumber\\
     &   \times \left\{\overline{\ell} X \ml(1 - \gama5 ){\nu}\right\},
\end{eqnarray}
where $m_D$ is the mass of the down-type quark ($s$, $b$) and $m_U$
is the mass of the up-type quark ($u$, $c$).  For comparison, the 
transition amplitude for the weak interaction is
\begin{equation}
  {\cal A}_W = -i\frac{G_F}{\sqrt{2}}V_{ij}
        \left\{\overline{U} \gamma_{\mu}(1 - \gama5 )D \right\}
        \left\{\overline{\ell} \gamma^{\mu}(1 - \gama5 )\nu\right\}.
\end{equation}
From the last two equations, one may read off the ratio of the 
Higgs coupling strength to the weak coupling strength.
\begin{equation}
   g_S = \frac{G_{H^{\pm}}}{G_{Weak}} =
              \frac{\ml m_D}{\mh^2}X^2\left( 1+\frac{m_U}{m_D} Y^2\right),
     \label{scalar_coupling_strength}
\end{equation}
where $m_D$ and $m_U$ are the current masses of the appropriate up and down
type quark, respectively and $|F_s/\fplus|=1$.  

The ratio, $g_S$, of the two couplings is the same $H^{\pm}$/$W^{\pm}$
coupling strength found in equation~\ref{ffratio}.  A measurement of 
the relative size of the scalar and vector form factors, then gives 
a relationship between the Higgs mass and the ratio of vacuum expectation 
values in type-II two Higgs doublet models:
\begin{equation}
   \mh = \left[\frac{\ml m_D}{g_s} 
         \left( 1+ \frac{m_U}{m_D X^2}\right)\right]^{\half} X.
\end{equation}

We now consider the application of the above phenomenology to $B$ meson 
decays.  

\section{$B$ meson decays}
In principle, one may fit the measured Dalitz plot of 
$B\rightarrow D\mu\nu$ decays for the relative admixture of the 
different form factors ($|\fminus/\fplus|$, $|\fs/\fplus|$, 
$|\ft/\fplus|$) and the relative phases of the form factors.
  
In practice,  the situation is more challenging. In the discussion
below, experience is derived from over two decades of kaon research
in the literature.  In $K\rightarrow\pi\ell\nu$ decays, the $Q^2$ 
dependence of the vector form factor has been shown to be 
linear~\cite{pdg:1998}, and is usually parameterized,
\begin{equation}
  \fplus(Q^2) = \fplus(0)\left( 1 + \lambda_{+}\frac{Q^2}{\mpi^2} \right).
\end{equation}
The values of $\lambda$ are small and may be approximated by 
a small change in slope of the phase space distribution in the $\Tpi$ 
direction.  In the $\Tpi$ direction, the effect of the $Q^2$ dependence
of the vector form factor is similar to introducing a small scalar 
coupling.  When fitting the Dalitz plot, $\lambda_+$ is correlated with 
$|\fs/\fplus|$.  As a further complication, electromagnetic 
radiative effects are significant in some regions of phase 
space~\cite{ke3gth,kmu3gth}.  If not taken into account, the net result 
produces a shift in the phase space distribution which would appear as an 
admixture of scalar and tensor couplings.  These effects are expected 
to manifest themselves in B decays in much the same manner as 
in kaon decays.

Semileptonic B decays of the form $B\rightarrow D\mu\nu$ appear to have
the most sensitivity to the effects described above.
In addition to the advantages of a larger coupling as demonstrated by
equation~\ref{scalar_coupling_strength}, using the muonic decays reduces 
electromagnetic radiative effects by approximately a factor of 
$m_{\mu}/m_{e}\approx 200$.  Unfortunately, from equation~\ref{dk_param}, 
the suppression of the induced scalar term ($\fminus$) for muonic decays 
is less than for electronic decays.

\begin{figure}[tbp]
%
%
 \epsfxsize=5.0in \epsfysize=5.0in
 \epsffile{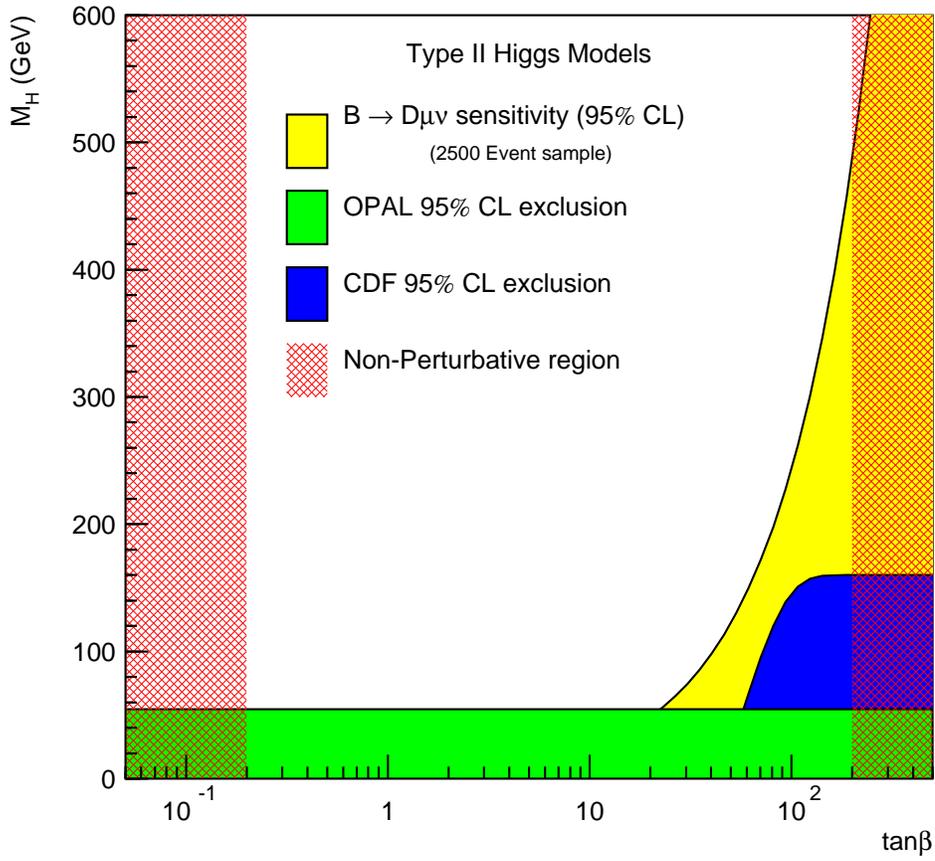}
 \caption{\label{b_higgs_limits} Charged Higgs 95\% 
  exclusion curves in the $(M_{H^{\pm}},\tan{\beta})$ plane for 
  $B\rightarrow D\mu\nu$ decays.  The CDF and OPAL results are 
  from direct searches.}
\end{figure}

In order to estimate the sensitivity of this method using B decay,
one may use the uncertainties from low statistics kaon results as a 
guide.  With a sample of 2500 exclusive $B\rightarrow D\mu\nu$ decays 
one would expect an induced scalar contribution of $\fminus = 0.02$
and an error on $|\fs/\fplus|\approx 0.04$.  Based on these numbers one
would expect an experimental sensitivity in the large $\tan{\beta}$ 
region of 
\begin{eqnarray}
  \mh < 2.44\tan{\beta}\;\rm GeV/c^2 \;(95\% CL).
\end{eqnarray}
In this calculation, we assume a $b$ quark mass of $4.4\;\rm GeV/c^2$.  
Figure~\ref{b_higgs_limits} compares the expected experimental charged
Higgs sensitivity in the $\mh$ vs $\tan{\beta}$ plane using the method
described above with current experimental results from direct 
searches~\cite{chhiggs:cdf,chhiggs:opal}.  
This method of indirect Higgs search shows significant sensitivity well 
beyond existing experimental data.  However, further increase of the 
sensitivity shown in the Figure by increasing the available statistics 
is likely not possible due to the induced scalar effects.  

Other indirect searches from the decays of $B$ mesons have set limits 
on the charged Higgs~\cite{CLEO:bsg,bxtau}.  The most restrictive 
of these searches comes from the CLEO measurement of 
BR($b\rightarrow s\gamma$) which gives an exclusion of 
$\mh > [244 + 63/(\tan{\beta})^{1.3}]$~GeV~\cite{CLEO:bsg}.  
Measuring the couplings as described above has sensitivity in the high,
$80 < \tan{\beta} < 200$ region.  Probably more important, since 
the above method measures the coupling to the hadronic current, 
penguin diagrams do not contribute to a scalar Lorentz structure.
Thus the above charged Higgs search technique can illuminate a 
portion of the charged Higgs parameter space not yet explored
in a way that is less model dependent.


\section{Summary}
Potential discovery of charged Higgs from direct production in 
$e^+e^-$ machines are limited by the beam energy (100 GeV for LEP).  
At the Tevatron, direct searches are currently limited by top 
production and the top mass~\cite{chhiggs:cdf}.  The direct 
searches are unlikely to exceed these limits in the near future.  
Therefore it becomes necessary to consider indirect methods for charged 
Higgs searches.

Using a traditional Dalitz plot analysis for $B$ meson decay,
we have shown sensitivity to the effects of charged Higgs.  Based on 
coarse estimates of the measurement uncertainties, we estimate that 
this technique is sensitive to charged Higgs in regions of the 
$\mh$ vs $\tan{\beta}$ parameter space previously unexplored.  Further, 
this method is sensitive to regions of parameter space inaccessible to 
direct searches in existing colliders.

The author acknowledges support from the National Science Foundation
under grant NSFPHY97-25210.  We also thank G. Farrar, N. Polonsky and 
S. Schnetzer for useful discussions and critical comments.

\end{document}